\documentclass[a4paper,12pt]{article}
\usepackage{color}
\usepackage{epsfig}
\usepackage{amsfonts}
\usepackage{amsmath}
\usepackage{amssymb}
\usepackage{rotating}
\usepackage{multirow}
\usepackage{mathtools}
\usepackage{chngpage}
\usepackage{setspace}
\usepackage{array}
\usepackage{url}
\usepackage[authoryear,comma,sectionbib]{natbib}

\begin{document}
\title{Mendelian randomization with a binary exposure variable: interpretation and presentation of causal estimates}
\author{Stephen Burgess \textsuperscript{1,2} \thanks{Corresponding author: Dr Stephen Burgess. Address: MRC Biostatistics Unit, Cambridge Institute of Public Health, Robinson Way, Cambridge, CB2 0SR, UK. Telephone: +44 1223 768259. Email: sb452@medschl.cam.ac.uk.} \and Jeremy A Labrecque \textsuperscript{3} \\ \\
\textsuperscript{1} MRC Biostatistics Unit, University of Cambridge, UK \\
\textsuperscript{2} Department of Public Health and Primary Care, \\ University of Cambridge, UK \\
\textsuperscript{3} Department of Epidemiology, Erasmus MC, Netherlands}
\maketitle

\noindent \noindent \textbf{Running head:} Mendelian randomization with binary exposure \\


\noindent \noindent \textbf{Acknowledgements:} The authors would like to thank Sonja A Swanson for her contribution to an earlier draft of this manuscript.

\maketitle

\clearpage
\setstretch{1.65}
\section*{Abstract}
Mendelian randomization uses genetic variants to make causal inferences about a modifiable exposure. Subject to a genetic variant satisfying the instrumental variable assumptions, an association between the variant and outcome implies a causal effect of the exposure on the outcome. Complications arise with a binary exposure that is a dichotomization of a continuous risk factor (for example, hypertension is a dichotomization of blood pressure). This can lead to violation of the exclusion restriction assumption: the genetic variant can influence the outcome via the continuous risk factor even if the binary exposure does not change. Provided the instrumental variable assumptions are satisfied for the underlying continuous risk factor, causal inferences for the binary exposure are valid for the continuous risk factor. Causal estimates for the binary exposure assume the causal effect is a stepwise function at the point of dichotomization. Even then, estimation requires further parametric assumptions. Under monotonicity, the causal estimate represents the average causal effect in `compliers', individuals for whom the binary exposure would be present if they have the genetic variant and absent otherwise. Unlike in randomized trials, genetic compliers are unlikely to be a large or representative subgroup of the population. Under homogeneity, the causal effect of the exposure on the outcome is assumed constant in all individuals; often an unrealistic assumption. We here provide methods for causal estimation with a binary exposure (although subject to all the above caveats). Mendelian randomization investigations with a dichotomized binary exposure should be conceptualized in terms of an underlying continuous variable.

\clearpage
\setstretch{1}

Mendelian randomization is the use of genetic variants as instrumental variables to test for or estimate the causal effect of a risk factor (referred to here as an exposure) on an outcome using observational data \citep{daveysmith2003, burgess2015book}. The primary objective of Mendelian randomization is to find modifiable exposures that are worthwhile therapeutic targets and can be intervened on to improve health outcomes. An instrumental variable must be associated with the exposure of interest (relevance), only affects the outcome through the exposure (exclusion restriction), and does not share any causes with the outcome (exchangeability). Recently, several Mendelian randomization studies have employed binary measures as the exposure variable. Examples include analyses assessing the causal effect of cannabis initiation on schizophrenia (and of schizophrenia on cannabis initiation) \citep{gage2016, vaucher2016}, and of diabetes status on endometrial cancer \citep{nead2015}. In this short manuscript, we discuss issues relating to causal estimation in the Mendelian randomization setting with a binary exposure. For ease of presentation, we initially assume a single genetic variant is used as an instrumental variable; this restriction is later relaxed.

The intended primary audience of this manuscript is Mendelian randomization practitioners, and the aim of the manuscript is to communicate the practical consequences of these methodological issues for Mendelian randomization investigations. As such, we focus on methods and approaches that are likely to be the most relevant to scenarios that are common in applied practice. In particular, we focus on methods that can be performed using summarized data, which comprise genetic associations with the exposure estimated using regression methods, that are routinely reported by large consortia \citep{burgess2013genepi}. Although our focus is on practitioners, we also provide technical asides and references for methodologically-focused readers.

\subsection*{Random assignment in a trial as a paradigm instrumental variable}
Consider a double-blind, placebo-controlled randomized trial with two time-fixed treatment arms (referred to as treatment and control) and complete follow-up data. An intention-to-treat effect estimate is typically reported: the causal effect of allocation to treatment as opposed to control. When there is substantial non-compliance, investigators may be interested in testing whether the treatment itself has an effect on the outcome (as opposed to simply allocation to treatment), or in estimating the causal effect of the treatment itself. Testing for a treatment or `per-protocol' effect can be achieved through the intention-to-treat analysis: unless random assignment somehow affects the outcome directly (e.g., because blinding is broken or a placebo effect is present), an association between treatment allocation and the outcome will only arise if the treatment has a causal effect on the outcome \citep{didelez2007}. Estimating the average treatment effect in the full study population further requires additional homogeneity conditions \citep{hernan2006, aronow2013, wang2018}; sufficient conditions are linearity of the instrumental variable--exposure, instrumental variable--outcome and exposure--outcome relationships with no effect heterogeneity. Without additional conditions, only bounds for the average treatment effect are obtainable \citep{balke1997}. These bounds can also be used to assess the validity of a genetic variant as an instrumental variable \citep{ramsahai2011, swanson2018}, although this approach is rarely informative in practice, and alternative ways of assessing instrument validity (such as understanding the biological role of the genetic variant, and assessing its associations with known confounders) are more likely to be fruitful in practice \citep{burgess2014twosample}. Alternatively, investigators often estimate an effect in a subgroup of the population under a weaker assumption.

Specifically, we consider the subgroup of the population consisting of `compliers' -- individuals who would receive the treatment if allocated to treatment, and would not receive treatment if allocated to not receive treatment. The effect in this subgroup can be estimated under the assumption that there are no defiers -- individuals who would only take treatment if randomly allocated not to do so, and who would not take treatment if allocated to take it \citep{angrist1996}. This is known as the monotonicity assumption -- allocation to taking the treatment can only increase the value of the exposure, not decrease it. This effect, which can be estimated using standard instrumental variable techniques, is known as the local average treatment effect (LATE) or the complier average causal effect (CACE) in the literature \citep{yau2001}. Of note, we cannot identify individual compliers as we cannot see individuals' treatment levels under both levels of treatment allocation. However, it is possible to identify the proportion of the study population who are compliers, and to describe relative characteristics of the compliers compared to non-compliers using measured baseline covariates \citep{angrist2009}. In well-designed randomized trials, compliers are likely to be common, and the assumption that there are no defiers is often considered reasonable.

\subsection*{Who are the genetic `compliers'?}
Monotonicity in the context of Mendelian randomization means that increasing the number of variant alleles for an individual can only increase the exposure from absent to present (or leave it constant), and can never decrease it. The analogue of `compliers' in Mendelian randomization are individuals who would have the exposure present if they possess an exposure-increasing genetic variant, but not otherwise. As genetic variants tend to have small effects on phenotypic variables, such compliers are likely to be uncommon.

This means that the group of genetic compliers is not likely to be representative of the general population. Also, the group of compliers may well differ greatly between different study populations. As an example, folate deficiency has been hypothesized as a causal risk factor for coronary heart disease \citep{lewis2005}. The complier population (and therefore the instrumental variable estimate) would differ greatly in a population where large numbers of people are borderline folate deficient compared with a population where relatively few people are folate deficient. (A similar problem would occur in randomized trials conducted in different populations.) The analogous assumption in Mendelian randomization to the `no defiers' assumption is that increases in the genotype variable would lead to increases (or no change) in the exposure for all individuals in the population (or equivalently, decreases or no change in the exposure for all individuals) \citep{hernan2006}. With a genetic variant that takes multiple values, the equivalent assumption is that the exposure is a non-decreasing (or non-increasing) function of the genetic variant. In this case (and in the case with multiple genetic variants), the instrumental variable estimate is a weighted average of LATEs \citep{angrist2000}.

In the context of RCTs, even if individual compliers cannot be identified, the subgroup of compliers may be of interest either because it represents a large or representative subgroup of the population, or due to patterns of non-compliance in the trial being anticipated to be repeated outside the trial setting. However, in Mendelian randomization, the subgroup of genetic `compliers' is unlikely to represent those individuals in the population who would respond to a treatment that influences the target exposure, particularly if the treatment has a greater effect on the risk factor than the genetic variant. Hence, under the `no defiers' assumption, the interpretation of a causal estimate in a Mendelian randomization investigation in which the instrumental variable assumptions are satisfied is that of an average causal effect in those individuals whose exposure status would vary depending on whether they have a particular genetic variant or not. We additionally note that the subgroup of genetic compliers would differ between genetic variants. This provides yet another reason why causal estimates based on different genetic variants may vary even if all the genetic variants are valid instruments. 

\subsection*{What is the true risk factor underlying the exposure?}
The above interpretation assumes that the instrumental variable assumptions are satisfied. These assumptions imply that the only influence of the instrumental variable on the outcome is via the exposure -- if the instrumental variable changes, but the exposure stays the same, then the outcome should not change. However, for most binary exposures used in Mendelian randomization investigations, there is an underlying continuous risk factor for which the binary variable is a dichotomization. As a simple example, the binary exposure hypertension is a dichotomization of the continuous risk factor blood pressure. In more complex examples, an underlying continuous latent variable can be hypothesized even if it cannot be measured, such as a continuous spectrum of sub-clinical mental health problems for the binary exposure schizophrenia.

If the binary exposure is a dichotomization of a continuous risk factor, then the instrumental variable assumptions are likely to be violated. For the example of hypertension, if elevated blood pressure is a causal risk factor for a particular outcome then genetic variants that are associated with blood pressure will be associated with the outcome even in a population where no-one suffers from clinically-defined hypertension. Hence, changes in the genetic variants will lead to increases in blood pressure and consequently to changes in the outcome even if the exposure status for hypertension remains fixed for all individuals in the population. An instrumental variable for a continuous exposure can only be an instrumental variable for the dichotomization of the exposure if the exposure--outcome causal relationship is a strict stepwise threshold at the point of dichotomization (in which case the dichotomized exposure is a representation of the true risk factor). However, provided that the instrumental variable assumptions are satisfied for the continuous risk factor, testing for an association with the outcome is still a valid test of the causal null hypothesis for the binary exposure.

There are two main consequences of this. First, such a Mendelian randomization study should be conceptualized as an investigation into the (possibly latent) underlying continuous risk factor, rather than the binary dichotomization of this variable. At minimum, the instrumental variable assumptions should be assessed with the continuous risk factor in mind. Second, a causal estimate from a Mendelian randomization investigation with a dichotomized binary exposure does not have a clear interpretation due to the binary exposure variable not capturing the true causal relationship. There are several reasons why a Mendelian randomization estimate may differ from the effect of an intervention even for a continuous exposure (for example, genetic variants have long-term influences acting from the beginning of life, whereas interventions are more short-term and are applied to mature individuals) \citep{burgess2012bmj, swanson2017}. With a binary exposure, these concerns are even greater. 

\subsection*{Causal estimation with a binary exposure}
Despite this, suppose that we want to calculate a causal effect with a binary exposure, under the assumption that the exposure has a stepwise effect on the outcome. This may be because we truly believe in the homogeneity assumptions, or we truly believe in the monotonicity assumption and regard the genetic compliers as a worthwhile subgroup of the population in which to estimate an average causal effect. Or, more likely, because a causal effect estimate is required for pragmatic reasons, such as to perform a power calculation or to inform policymakers of the expected impact of intervention on the exposure. Other reasons for estimating a causal parameter include efficient testing of the causal null hypothesis with multiple instrumental variables (under the homogeneity assumptions, the two-stage least squares estimate, or equivalently the inverse-variance weighted estimate, is the optimally efficient combination of the instruments for testing for a causal effect \citep{wooldridge2009ch15}) and use of a robust method with multiple genetic variants (such as the MR-Egger method \citep{bowden2015} or weighted median method \citep{bowden2015median} -- these methods make weaker assumptions, not requiring all genetic variants to satisfy the instrumental variable assumptions). If the binary exposure is a dichotomization of a continuous risk factor, then power calculations are likely to be conservative, as the effect of the genetic variant on the outcome will not be fully captured by the binary exposure.

Two options for causal estimation are: i) estimating the effect on the outcome per (say) 1\% absolute increase in the probability of the exposure; ii) estimating the effect on the outcome per (say) doubling of the probability (or odds) of the exposure. We concentrate on estimation methods based on regression (usually linear or logistic) for several reasons. First, often researchers perform their analyses using summarized association estimates -- beta-coefficients from regression analyses of the exposure and outcome on a genetic variant -- and do not have access to individual-level data. These beta-coefficients represent the average change in the trait (exposure or outcome) per additional copy of the effect allele. Secondly, these approaches result in causal estimates with a simple and relevant interpretation, and which can be compared to estimates in the literature from other analytical approaches. Thirdly, often there are technical restrictions on the data analysis -- for example, it may be necessary to fit a mixed model to account for relatedness between individuals, to adjust for several principal components of ancestry, or to provide a coordinated approach to analysis across different datasets. These restrictions are easiest to accommodate in a regression framework. These estimation procedures require strict linearity and homogeneity assumptions; full details are available elsewhere \citep{hernan2006, didelez2007}. The parametric assumptions for these two options are mutually incompatible. Additionally, regression coefficients will generally be variation dependent on the baseline risk, a nuisance parameter \citep{richardson2017}. If individual-level data are available, then alternative approaches to estimation can be taken \citep{aronow2013, wang2018}.

If the genetic associations with the exposure are estimated using linear regression, then they represent absolute changes in the prevalence of the exposure. This enables estimation of the causal effect of an intervention in the prevalence of the exposure on an absolute scale. It is sensible to scale the causal effect to consider a modest increase in the prevalence of the exposure (say a 1\% or a 10\% increase), as a unit increase would represent the average causal effect of a population intervention from 0\% prevalence of the exposure to 100\% prevalence -- an unrealistic intervention in practice. However, absolute associations with a binary variable do not make sense in case-control settings (where cases are those with the exposure), as they depend on the ratio of cases to controls chosen by the investigator.

If the genetic associations with the exposure are estimated using logistic regression, then they represent log odds ratios. The causal estimate would then represent the change in the outcome per unit change in the exposure on the log odds scale. A unit increase in the log odds of a variable corresponds to a 2.72 ($= \exp 1$)-fold multiplicative increase in the odds of the variable. If the exposure is rare then the odds of the exposure is approximately equal to the probability of the exposure. The causal estimate represents the average change in the outcome per 2.72-fold increase in the prevalence of the exposure (for example, an increase in the exposure prevalence from 1\% to 2.72\%). It may be more interpretable to think instead about the average change in the outcome per doubling (2-fold increase) in the prevalence of the exposure. This can be obtained by multiplying the causal estimate by 0.693 ($= \log_e 2$).

\subsection*{Discussion}
\begin{figure}[htp]
\centering
\includegraphics[width=1.1 \textwidth, clip, bb = 0 0 636 474]{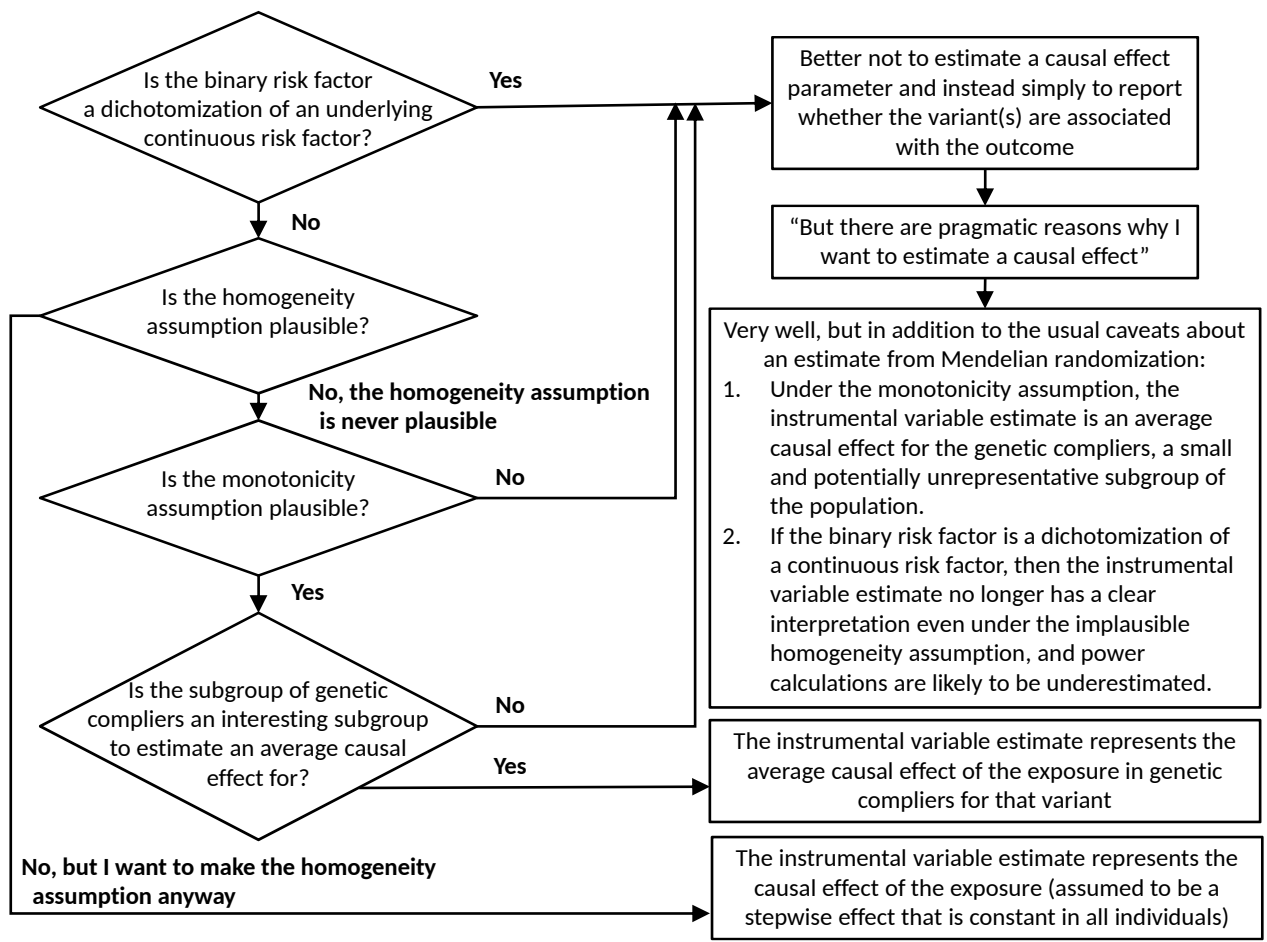}
\caption{Flow diagram illustrating the steps needed to consider when considering whether to estimate a parameter in a Mendelian randomization investigation or not with a binary risk factor.} \label{summ}
\end{figure}

In this short manuscript, we have discussed statistical issues for Mendelian randomization with a binary exposure. A summary of the arguments made in the paper is provided as Figure~\ref{summ}. Under the more plausible assumption of monotonicity, the estimate from a Mendelian randomization study with a binary exposure represents the average causal effect in `compliers'; the subgroup of individuals for whom the presence or absence of the genetic variant used as an instrument determines whether individuals have the exposure present or not. Under the less plausible assumption of homogeneity, the estimate of the causal effect only makes sense if the effect of the exposure on the outcome has a strict stepwise form -- only changes in whether the binary exposure is present or absent will affect the outcome. If the binary exposure is a dichotomization of a continuous variable, then the causal estimate does not have a clear interpretation. In such a case, causal inferences will only be valid provided that the instrumental variable assumptions are satisfied for the continuous risk factor -- in particular, if the effect of the genetic variant on the outcome is completely mediated via the continuous risk factor. However, as the effect of the genetic variant on the outcome is not completely mediated via the binary exposure, power calculations are likely to be conservative.

In summary, applying Mendelian randomization with a binary exposure requires careful consideration. When the binary exposure is a dichotomization of an underlying continuous risk factor, causal assumptions should be assessed and causal inferences should be conceptualized with respect to the underlying continuous risk factor. Tests for causal effects may be achieved readily without using the exposure information, but estimation procedures for a binary exposure require strong assumptions that are unlikely to be biologically plausible in common Mendelian randomization settings.

\vspace{6mm}

\noindent \noindent \textbf{Funding:} Stephen Burgess is supported by a Sir Henry Dale Fellowship jointly funded by the Wellcome Trust and the Royal Society (Grant Number 204623/Z/16/Z). \\

\noindent \noindent \textbf{Conflict of Interest:} The authors declare that they have no conflict of interest.

\bibliographystyle{DeGruyter} 
\bibliography{masterref_sonja}

\begin{thebibliography}{27}
\newcommand{\enquote}[1]{``#1''}
\providecommand{\natexlab}[1]{#1}
\providecommand{\url}[1]{\texttt{#1}}
\providecommand{\urlprefix}{URL }

\bibitem[{Angrist et~al.(2000)Angrist, Graddy, and Imbens}]{angrist2000}
Angrist, J., K.~Graddy, and G.~Imbens (2000): \enquote{{The interpretation of
  instrumental variables estimators in simultaneous equations models with an
  application to the demand for fish},} \emph{Review of Economic Studies}, 67,
  499--527.

\bibitem[{Angrist et~al.(1996)Angrist, Imbens, and Rubin}]{angrist1996}
Angrist, J., G.~Imbens, and D.~Rubin (1996): \enquote{{Identification of causal
  effects using instrumental variables},} \emph{Journal of the American
  Statistical Association}, 91, 444--455.

\bibitem[{Angrist and Pischke(2009)}]{angrist2009}
Angrist, J. and J.~Pischke (2009): \emph{{Mostly harmless econometrics: an
  empiricist's companion. Chapter 4: Instrumental variables in action:
  sometimes you get what you need}}, Princeton University Press.

\bibitem[{Aronow and Carnegie(2013)}]{aronow2013}
Aronow, P.~M. and A.~Carnegie (2013): \enquote{{Beyond LATE: Estimation of the
  average treatment effect with an instrumental variable},} \emph{Political
  Analysis}, 21, 492--506.

\bibitem[{Balke and Pearl(1997)}]{balke1997}
Balke, A. and J.~Pearl (1997): \enquote{{Bounds on treatment effects from
  studies with imperfect compliance},} \emph{Journal of the American
  Statistical Association}, 92, 1171--1176.

\bibitem[{Bowden et~al.(2015)Bowden, Davey~Smith, and Burgess}]{bowden2015}
Bowden, J., G.~Davey~Smith, and S.~Burgess (2015): \enquote{{Mendelian
  randomization with invalid instruments: effect estimation and bias detection
  through Egger regression},} \emph{International Journal of Epidemiology}, 44,
  512--525.

\bibitem[{Bowden et~al.(2016)Bowden, Davey~Smith, Haycock, and
  Burgess}]{bowden2015median}
Bowden, J., G.~Davey~Smith, P.~C. Haycock, and S.~Burgess (2016):
  \enquote{{Consistent estimation in Mendelian randomization with some invalid
  instruments using a weighted median estimator},} \emph{Genetic Epidemiology},
  40, 304--314.

\bibitem[{Brion et~al.(2013)Brion, Shakhbazov, and Visscher}]{brion2013}
Brion, M.-J., K.~Shakhbazov, and P.~Visscher (2013): \enquote{{Calculating
  statistical power in Mendelian randomization studies},} \emph{International
  Journal of Epidemiology}, 42, 1497--1501.

\bibitem[{Burgess(2014)}]{burgess2013power}
Burgess, S. (2014): \enquote{{Sample size and power calculations in Mendelian
  randomization with a single instrumental variable and a binary outcome},}
  \emph{International Journal of Epidemiology}, 43, 922--929.

\bibitem[{Burgess et~al.(2012)Burgess, Butterworth, Malarstig, and
  Thompson}]{burgess2012bmj}
Burgess, S., A.~Butterworth, A.~Malarstig, and S.~Thompson (2012):
  \enquote{{Use of Mendelian randomisation to assess potential benefit of
  clinical intervention},} \emph{British Medical Journal}, 345, e7325.

\bibitem[{Burgess et~al.(2013)Burgess, Butterworth, and
  Thompson}]{burgess2013genepi}
Burgess, S., A.~S. Butterworth, and S.~G. Thompson (2013): \enquote{{Mendelian
  randomization analysis with multiple genetic variants using summarized
  data},} \emph{Genetic Epidemiology}, 37, 658--665.

\bibitem[{Burgess et~al.(2015)Burgess, Scott, Timpson, Davey~Smith, Thompson,
  and {EPIC-InterAct Consortium}}]{burgess2014twosample}
Burgess, S., R.~Scott, N.~Timpson, G.~Davey~Smith, S.~G. Thompson, and
  {EPIC-InterAct Consortium} (2015): \enquote{{Using published data in
  Mendelian randomization: a blueprint for efficient identification of causal
  risk factors},} \emph{European Journal of Epidemiology}, 30, 543--552.

\bibitem[{Burgess and Thompson(2015)}]{burgess2015book}
Burgess, S. and S.~G. Thompson (2015): \emph{{Mendelian randomization: methods
  for using genetic variants in causal estimation}}, Chapman \& Hall.

\bibitem[{Davey~Smith and Ebrahim(2003)}]{daveysmith2003}
Davey~Smith, G. and S.~Ebrahim (2003): \enquote{{`Mendelian randomization': can
  genetic epidemiology contribute to understanding environmental determinants
  of disease?}} \emph{International Journal of Epidemiology}, 32, 1--22.

\bibitem[{Didelez and Sheehan(2007)}]{didelez2007}
Didelez, V. and N.~Sheehan (2007): \enquote{{Mendelian randomization as an
  instrumental variable approach to causal inference},} \emph{Statistical
  Methods in Medical Research}, 16, 309--330.

\bibitem[{Gage et~al.(2017)Gage, Jones, Burgess, Bowden, Smith, Zammit, and
  Munaf{\`o}}]{gage2016}
Gage, S.~H., H.~J. Jones, S.~Burgess, J.~Bowden, G.~D. Smith, S.~Zammit, and
  M.~R. Munaf{\`o} (2017): \enquote{{Assessing causality in associations
  between cannabis use and schizophrenia risk: a two-sample Mendelian
  randomization study},} \emph{Psychological Medicine}.

\bibitem[{Hern{\'a}n and Robins(2006)}]{hernan2006}
Hern{\'a}n, M. and J.~Robins (2006): \enquote{{Instruments for causal
  inference: an epidemiologist's dream?}} \emph{Epidemiology}, 17, 360--372.

\bibitem[{Lewis et~al.(2005)Lewis, Ebrahim, and Davey~Smith}]{lewis2005}
Lewis, S., S.~Ebrahim, and G.~Davey~Smith (2005): \enquote{{Meta-analysis of
  MTHFR 677C - T polymorphism and coronary heart disease: does totality of
  evidence support causal role for homocysteine and preventive potential of
  folate?}} \emph{British Medical Journal}, 331, 1053.

\bibitem[{Nead et~al.(2015)Nead, Sharp, Thompson, Painter, Savage, Semple,
  Barker, Perry, Attia, Dunning et~al.}]{nead2015}
Nead, K.~T., S.~J. Sharp, D.~J. Thompson, J.~N. Painter, D.~B. Savage, R.~K.
  Semple, A.~Barker, J.~R. Perry, J.~Attia, A.~M. Dunning, et~al. (2015):
  \enquote{{Evidence of a causal association between insulinemia and
  endometrial cancer: a Mendelian randomization analysis},} \emph{Journal of
  the National Cancer Institute}, 107, djv178.

\bibitem[{Ramsahai and Lauritzen(2011)}]{ramsahai2011}
Ramsahai, R. and S.~Lauritzen (2011): \enquote{{Likelihood analysis of the
  binary instrumental variable model},} \emph{Biometrika}, 98, 987--994.

\bibitem[{Richardson et~al.(2017)Richardson, Robins, and Wang}]{richardson2017}
Richardson, T.~S., J.~M. Robins, and L.~Wang (2017): \enquote{{On modeling and
  estimation for the relative risk and risk difference},} \emph{Journal of the
  American Statistical Association}, 112, 1121--1130.

\bibitem[{Swanson et~al.(2018)}]{swanson2018}
Swanson, S. et~al. (2018): \enquote{Partial identification of the average
  treatment effect using instrumental variables: review of methods for binary
  instruments, treatments, and outcomes,} \emph{Journal of the American
  Statistical Association}.

\bibitem[{Swanson et~al.(2017)Swanson, Tiemeier, Ikram, and
  Hern{\'a}n}]{swanson2017}
Swanson, S.~A., H.~Tiemeier, M.~A. Ikram, and M.~A. Hern{\'a}n (2017):
  \enquote{{Nature as a trialist?: Deconstructing the analogy between Mendelian
  randomization and randomized trials},} \emph{Epidemiology}, 28, 653--659.

\bibitem[{Vaucher et~al.(2017)Vaucher, Keating, Lasserre, Gan, Lyall, Ward,
  Smith, Pell, Sattar, Pare, and Holmes}]{vaucher2016}
Vaucher, J., B.~J. Keating, A.~M. Lasserre, W.~Gan, D.~Lyall, J.~Ward, D.~J.
  Smith, J.~Pell, N.~Sattar, G.~Pare, and M.~Holmes (2017): \enquote{Cannabis
  use and risk of schizophrenia: a mendelian randomization study,}
  \emph{Molecular Psychiatry}.

\bibitem[{Wang and Tchetgen~Tchetgen(2018)}]{wang2018}
Wang, L. and E.~Tchetgen~Tchetgen (2018): \enquote{{Bounded, efficient and
  triply robust estimation of average treatment effects using instrumental
  variables},} \emph{arXiv}, 1611.09925.

\bibitem[{Wooldridge(2009)}]{wooldridge2009ch15}
Wooldridge, J. (2009): \emph{{Introductory econometrics: A modern approach.
  Chapter 15: Instrumental variables estimation and two stage least squares}},
  South-Western, Nashville, TN.

\bibitem[{Yau and Little(2001)}]{yau2001}
Yau, L.~H. and R.~J. Little (2001): \enquote{{Inference for the
  complier-average causal effect from longitudinal data subject to
  noncompliance and missing data, with application to a job training assessment
  for the unemployed},} \emph{Journal of the American Statistical Association},
  96, 1232--1244.

\end{thebibliography}

\clearpage

\end{document}